\documentclass[10pt,conference]{IEEEtran}
\IEEEoverridecommandlockouts
\usepackage{cite}
\usepackage{amsmath,amssymb,amsfonts}
\usepackage{algorithmic}
\usepackage{graphicx}
\usepackage{textcomp}
\usepackage{cite}
\usepackage{amsmath,amssymb,amsfonts}
\usepackage{comment}
\usepackage{algorithmic}
\usepackage{graphicx}
\usepackage{xcolor}
\usepackage{orcidlink}
\hypersetup{hidelinks}
\usepackage{pdflscape}
\usepackage[caption=false,font=footnotesize]{subfig}
\usepackage{tabularx}
\usepackage{longtable}
\usepackage{rotating}
\usepackage{lipsum} % For dummy text
\usepackage{adjustbox}
\usepackage{textcomp}
\usepackage{geometry}
\usepackage{listings}
\lstdefinestyle{configstyle}{
    basicstyle=\ttfamily\footnotesize,
    backgroundcolor=\color{gray!10},
    frame=single,
    breaklines=true,
    columns=fullflexible
}
\usepackage{balance}

\usepackage{booktabs}%,caption}
\usepackage[flushleft]{threeparttable}
\usepackage{xcolor}
\def\BibTeX{{\rm B\kern-.05em{\sc i\kern-.025em b}\kern-.08em
    T\kern-.1667em\lower.7ex\hbox{E}\kern-.125emX}}

\usepackage{acro}
\usepackage{array}
 \usepackage{amsmath}
\usepackage{amssymb}% http://ctan.org/pkg/amssymb
\usepackage{pifont}% http://ctan.org/pkg/pifont

\usepackage{graphicx}
\usepackage{tikz}
\usepackage{comment}
\usepackage{textcomp}
\usepackage{hyperref} 

\DeclareAcronym{3G}{short=3G, long=third generation}
\DeclareAcronym{3GPP}{short=3GPP, long=Third Generation Partnership Project}
\DeclareAcronym{4G}{short=4G, long=fourth generation}
\DeclareAcronym{5G}{short=5G, long=fifth generation}
\DeclareAcronym{6G}{short=6G, long=sixth generation}

\DeclareAcronym{ASiR}{short = ASiR , long = AirScale indoor radiohead}
\DeclareAcronym{AI}{short=AI, long=artificial intelligence}
\DeclareAcronym{AMF}{short=AMF, long=access and mobility management functions}

\DeclareAcronym{BBU}{short = BBU ,  long = baseband unit}

\DeclareAcronym{CA}{short = CA , long =  certificate authority}
\DeclareAcronym{CN}{short = CN , long = core network}
\DeclareAcronym{COTS}{short = COTS , long = commercial Off-The-Shelf}
\DeclareAcronym{CU}{short = CU , long = central unit}
\DeclareAcronym{CP}{short = CP , long = control plane}
\DeclareAcronym{C-RAN}{short = C-RAN , long = cloud radio access network}
\DeclareAcronym{CPU}{short = CPU , long = Central Processing Unit}
 
\DeclareAcronym{DAS}{short = DAS , long = distributed antenna system}
\DeclareAcronym{DDoS}{short = DDoS, long = distributed denial-of-service}
\DeclareAcronym{DoS}{short = DoS , long = denial-of-service}
\DeclareAcronym{DL}{short = DL, long = downlink}
\DeclareAcronym{DU}{short = DU , long = Distributed unit}
\DeclareAcronym{DPI}{short = DPI , long = deep packet inspection}
\DeclareAcronym{DRS}{short = DRS , long = distributed radio system}
\DeclareAcronym{DN}{short = DN , long = data network}
\DeclareAcronym{DTLS}{short = DTLS, long = Datagram Transport Layer Security}

\DeclareAcronym{E2SM}{short = E2SM , long = E2 service model}
\DeclareAcronym{E2E}{short = E2E , long = End-to-End}
\DeclareAcronym{EIRP}{short = EIRP , long = effective isotropic radiated power}
\DeclareAcronym{eMBB}{short = eMBB , long = enhanced mobile broadband}
\DeclareAcronym{EMF}{short = EMF , long = electromagnetic field}
\DeclareAcronym{ESIM}{short = ESIM , long = earth stations in motion}
\DeclareAcronym{ETSI}{short = ETSI , long = European Telecommunications Standards Institute}
\DeclareAcronym{EU}{short = EU , long = European Union}
\DeclareAcronym{ESP}{short=ESP, long=encapsulating security payload}
\DeclareAcronym{E2AP}{short = E2AP , long = E2 Application Protocol}

\DeclareAcronym{FCC}{short = FCC , long = Federal Communications Commission}
\DeclareAcronym{FDD}{short = FDD , long = frequency division duplex}
\DeclareAcronym{FR1}{short = FR1 , long = frequency range 1}
\DeclareAcronym{FR2}{short = FR2 , long = frequency range 2}
\DeclareAcronym{FSS}{short = FSS , long = fixed satellite service}

\DeclareAcronym{GEO}{short = GEO , long = geostationary earth orbit}
\DeclareAcronym{gNB}{short = gNB , long = next generation NodeB}
\DeclareAcronym{GPU}{short = GPU , long =graphics processing unit}

\DeclareAcronym{HAPS}{short = HAPS , long = high altitude platform station}

\DeclareAcronym{ICES}{short = ICES , long = International Committee on Electromagnetic Safety}
\DeclareAcronym{IKEv2}{short =IKEv2 , long = Internet Key Exchange version 2}
\DeclareAcronym{ICNIRP}{short = ICNIRP , long = International Commission on Non-Ionizing Radiation Protection}
\DeclareAcronym{IEC}{short = IEC , long = International Electrotechnical Commission}
\DeclareAcronym{IEEE}{short = IEEE , long = Institute of Electrical and Electronics Engineers}
\DeclareAcronym{IP}{short = IP , long = Internet Protocol}
\DeclareAcronym{IMT}{short = IMT , long = International Mobile Telecommunications}
\DeclareAcronym{IoT}{short = IoT , long = Internet of Things}
\DeclareAcronym{IIoT}{short = IIoT , long = industrial internet of things}
\DeclareAcronym{IPsec}{short = IPsec , long = internet protocol security}
\DeclareAcronym{ITU}{short = ITU , long = International Telecommunication Union}
\DeclareAcronym{ICMP}{ short = ICMP , long =internet control message protocol}

\DeclareAcronym{KPI}{short = KPI , long = key performance indicators}
\DeclareAcronym{LEO}{short = LEO , long = low earth orbit}
\DeclareAcronym{LOS}{short = LOS , long = line of sight}
\DeclareAcronym{LTE}{short = LTE , long = long term evolution}

\DeclareAcronym{MAC}{short = MAC , long = message authentication codes}
\DeclareAcronym{MEO}{short = MEO , long = medium earth orbit}
\DeclareAcronym{MitM}{short = MitM , long = man-in-the-middle}
\DeclareAcronym{ML}{short = ML , long = Machine Learning}
\DeclareAcronym{mMIMO}{short = mMIMO , long = massive multiple input multiple output}
\DeclareAcronym{mMTC}{short = mMTC , long = massive machine type communication}

\DeclareAcronym{mmWave}{short = mmWave , long = millimeter wave}
\DeclareAcronym{MPC}{short = MPC , long =multipath components}
\DeclareAcronym{mRRH}{short = mRRH , long = micro-RRH}
\DeclareAcronym{MSS}{short = MSS , long = mobile satellite service}

\DeclareAcronym{NAS}{short = NAS , long = e Non-Access Stratum}
\DeclareAcronym{NATO}{short = NATO , long = North Atlantic Treaty Organization}
\DeclareAcronym{NGAP}{short = NGAP , long = Next Generation Application Protocol}
\DeclareAcronym{NCRP}{short = NCRP , long =national council on radiation protection and measurements}
\DeclareAcronym{NDAC}{short = NDAC , long = Nokia Digital Automation Cloud}
\DeclareAcronym{NG-RAN}{short = NG-RAN , long = next generation radio access network}
\DeclareAcronym{NGSO}{short = NGSO , long = non-geostationary satellite orbit}
\DeclareAcronym{NR}{short = NR , long = new radio}
\DeclareAcronym{NSA}{short = NSA , long = non-stand-alone}
\DeclareAcronym{NTN}{short = NTN , long = non-terrestrial network}

\DeclareAcronym{OAI}{short = OAI , long = OpenAirInterface}
\DeclareAcronym{O-FH}{short = O-FH , long = open fronthaul}
\DeclareAcronym{O-RAN}{short = O-RAN , long = open radio access network}
\DeclareAcronym{O-CU}{short = O-CU , long = open central unit}
\DeclareAcronym{O-DU}{short = O-DU , long = open distributed unit}
\DeclareAcronym{OT}{short = OT, long = Operational Technology}

\DeclareAcronym{PBCH}{short = PBCH , long = physical broadcast channel}
\DeclareAcronym{PCI}{short = PCI , long = physical cell ID}
\DeclareAcronym{PDCP}{short = PDCP , long = packet data convergence protocol}
\DeclareAcronym{PDSCH}{short = PDSCH , long = physical downlink shared channel}
\DeclareAcronym{PLMN}{short = PLMN , long = public land mobile network}
\DeclareAcronym{PoE}{short = PoE , long = power over ethernet}
\DeclareAcronym{pRRH}{short = pRRH , long = pico-RRH}
\DeclareAcronym{PTP}{short = PTP , long =precision time protocol}

\DeclareAcronym{QoS}{short = QoS , long = quality of service}
\DeclareAcronym{RAN}{short = RAN , long = radio access network}
\DeclareAcronym{RAT}{short = RAT , long = radio access technology}
\DeclareAcronym{RF}{short = RF , long = radio frequency}
\DeclareAcronym{RIC}{short = RIC , long = RAN Intelligent Controller}
\DeclareAcronym{RRH}{short = RRH , long = remote radio heads}
\DeclareAcronym{RSS}{short = RSS , long = root sum square}
\DeclareAcronym{RSRP}{short = RSRP, long = reference signal received power}
\DeclareAcronym{RSSI}{short = RSSI, long = received signal strength indication}
\DeclareAcronym{RT}{short = RT , long = Real-time}
\DeclareAcronym{RTT}{short=RTT, long=round-trip time}
\DeclareAcronym{RU}{short = RU , long = radio unit}

\DeclareAcronym{SCTP}{short = SCTP , long = Stream Control Transmission Protocol}
\DeclareAcronym{SA}{short = SA , long = Security Associations}
\DeclareAcronym{SAN}{short = SAN ,  long = satellite access node}
\DeclareAcronym{SAR}{short = SAR , long = specific absorption rate}
\DeclareAcronym{SINR}{short = SINR , long = Signal-to-interference-plus-noise ratio ratio }
\DeclareAcronym{SMO}{short = SMO , long = service management orchestration}
\DeclareAcronym{SSB}{short = SSB , long = synchronization signal block}

\DeclareAcronym{TDD}{short = TDD , long = time division duplex}
\DeclareAcronym{TDP}{short = TDP , long = thermal design power}
\DeclareAcronym{TDR}{short = TDR , long = thermal design ratio}
\DeclareAcronym{TEID}{short = TEID , long = Tunnel Endpoint Identifier}
\DeclareAcronym{THz}{short = THz , long = terahertz}
\DeclareAcronym{TN}{short = TN , long = terrestrial network}
\DeclareAcronym{TNL}{short = TNL , long = transport network layer}
\DeclareAcronym{TUK}{short = TUK, long = Technische Universit\"at Kaiserslautern}
\DeclareAcronym{TLS}{short = TLS, long = Transport Layer Security}

\DeclareAcronym{UAS}{short = UAS , long = unmanned aerial system }
\DeclareAcronym{UE}{short = UE , long = user equipment}
\DeclareAcronym{UP}{short = UP , long = user plane}
\DeclareAcronym{UPF}{short = UPF , long = user plane function}
\DeclareAcronym{UDP}{short = UDP , long =user datagram protocol}

\DeclareAcronym{VSAT}{short = VSAT , long = very small aperture terminal }
  \DeclareAcronym{V-RAN}{short = V-RAN , long = virtual radio access network}
  \DeclareAcronym{VPN}{short = VPN , long = Virtual Private Network}

\DeclareAcronym{XR}{short = XR , long = extended reality}
\DeclareAcronym{XRF}{short = XRF , long = xApp repository function }

\DeclareAcronym{WHO}{short = WHO , long =world health organization}

\DeclareAcronym{ZTA}{short = ZTA , long = zero trust architecture}

\usepackage{multirow}
\usepackage{float}
\usepackage{color,soul}
\usepackage{xurl}

\usepackage{xcolor}
\def\BibTeX{{\rm B\kern-.05em{\sc i\kern-.025em b}\kern-.08em
    T\kern-.1667em\lower.7ex\hbox{E}\kern-.125emX}}

\makeatother
\usepackage{eso-pic}
\newcommand\AtPageUpperMyright[1]{\AtPageUpperLeft{
\put(\LenToUnit{0.08\paperwidth},\LenToUnit{-1cm}){
    \parbox{1\textwidth}{\fontsize{8}{11}\selectfont #1}}
}}
\newcommand{\conf}[1]{
\AddToShipoutPictureBG*{
\AtPageUpperMyright{#1}
}
}
\begin{document}

%To print copyright header
\conf{This paper has been submitted to IEEE. Copyright may be transferred without notice, so this version might differ from published version.}
\vspace{-3mm}
%%%%%%%%%%%%%%%%%%%%

%\title{Towards WireGuard-Enabled Lightweight Security for Industrial 6G Open RAN}
\title{Lightweight Security for Private Networks: Real-World Evaluation of WireGuard}
\vspace{-4.5mm}
%\title{Wireguard for lightweight security in 6G industrial Open RAN Campus networks
%\title{Lightweight Security for Industrial Open RAN: Evaluating WireGuard Against IPsec in a Real-World Factory Deployment}

%\thanks{Received on \textcolor{blue}{DD MM 2025}; revised on \textcolor{blue}{DD MM 2025}; and accepted for publication by \textcolor{blue}{Associate Editor} on \textcolor{blue}{DD MM 2025}. Date of publication \textcolor{blue}{DD MM 2025}; date of current version \textcolor{blue}{DD MM 2025}.}
%\thanks{Hubert Djuitcheu and Andrew Sergeev are with Adtran Networks SE, Meiningen, Germany}
%\thanks{Jochen Seitz is with Communication Networks Group, TU Ilmenau, Germany}
%\thanks{This work was funded in part by...}

\author{
\IEEEauthorblockN{
Hubert Djuitcheu\orcidlink{0000-0002-2650-8780}\textsuperscript{$\diamond$}, 
Andrew Sergeev\textsuperscript{$\diamond$}, 
Khurshid Alam\orcidlink{0009-0003-3344-8629}\textsuperscript{$*$}, 
Danny Santhosh\textsuperscript{$\diamond$},
Achim Autenrieth\orcidlink{0000-0002-3475-2038}\textsuperscript{$\diamond$}, \\
Jochen Seitz\orcidlink{0000-0001-7867-9680}\textsuperscript{$\beta$}
}
\IEEEauthorblockA{ 
\textsuperscript{$\diamond$}\textit{Adtran Networks SE,} Meiningen, Germany.\\ 
 \{hubert.djuitcheu, andrew.sergeev, danny.santhosh, achim.autenrieth\}@adtran.com \\
\textsuperscript{$*$}\textit{German Research Center for Artificial Intelligence (DFKI),} Kaiserslautern, Germany.\\
	khurshid.alam@dfki.de\\
\textsuperscript{$\beta$}\textit{Communication Networks Group,} TU Ilmenau, Germany.\\ 
jochen.seitz@tu-ilmenau.de\\
}
}
\vspace{-3mm}
\maketitle
\vspace{-3.5mm}

\begin{abstract}
This paper explores WireGuard as a lightweight alternative to IPsec for securing the user plane as well as the control plane in an industrial Open RAN deployment at the Adtran Terafactory in Meiningen. We focus on a realistic scenario where external vendors access their hardware in our 5G factory network, posing recurrent security risks from untrusted gNBs and intermediate network elements. Unlike prior studies limited to lab setups, we implement a complete proof-of-concept in a factory environment and compare WireGuard with IPsec under industrial traffic conditions. Our approach successfully protects user data (N3 interface) against untrusted gNBs and man-in-the-middle attacks while enabling control plane (N2 interface) authentication between the \ac{AMF} and gNB. Performance measurements show that WireGuard adds minimal overhead in throughput, latency, and \ac{CPU} usage, achieving performance comparable to IPsec. These findings demonstrate that WireGuard offers competitive performance with significantly reduced configuration complexity, making it a strong candidate for broader adoption in O-RAN, providing a unified, lightweight security layer across multiple interfaces and components.

%The adoption of sixth-generation (6G) networks and open \ac{RAN} paradigms in industrial campus network deployments enables vendor flexibility and innovation, but also introduces new security challenges, including exposed fronthaul interfaces, untrusted software components, and increased attack surfaces. In this paper, we investigate the use of the WireGuard VPN as a lightweight, unified security layer across both the \ac{CP} and \ac{UP} of an industrial open \ac{RAN} campus network. Our contributions are fourfold: (i) a threat analysis targeting rogue cells, supply-chain compromises, and misconfigurations; (ii) a deployment design that integrates WireGuard tunnels between the next-generation NodeBs (gNBs) (\ac{OAI})-based open-source and commercial Synergy gNBs) and \ac{CN} functions (\ac{UPF}, \ac{AMF}) within the \emph{6G-Terafactory} testbed; (iii) an extension concept leveraging Merkle-tree based attestation for CP integrity; and (iv) a comprehensive evaluation of security, latency, and operational overhead. Experimental results demonstrate that WireGuard’s minimal cryptographic codebase (kernel-mode ChaCha20-Poly1305 with ECDH key exchange) incurs negligible latency and CPU cost, while mitigating eavesdropping, spoofing, and \ac{MitM} attacks. Compared to IPsec, WireGuard achieves faster deployment, lower complexity, and strong automation support. These findings indicate that WireGuard is a practical candidate for lightweight security in industrial 6G open \ac{RAN} campus networks.
\end{abstract}

%\IEEEoverridecommandlockouts

%%%%%%%%%%%%%%%%%%%%%%%%%%%  document   %%%%%%%%%%%%%%%%%%%%%%%%%%%%%%%%%%%%%%%%
\vspace{-2.5mm}
%%%%%%%%%%%%%%%%%%%%%%%%%%%%%%%%%%%    Section Introduction   %%%%%%%%%%%%%%%%%%%%%%%%%%%%%%%%%%%%%%

\section{Introduction}

Open RAN is gaining traction for private and industrial 5G networks because it enables multi-vendor deployments, faster innovation, and flexible integration with existing \ac{OT} infrastructures \cite{11124199}. However, the increased openness and disaggregation of the radio access network also enlarge the attack surface, exposing interfaces, software stacks, and supply chains that were previously confined to tightly controlled vendor domains. In industrial campuses where external vendors require remote access to machines and production assets, these weaknesses are amplified by the presence of untrusted gNBs, shared backhaul segments, and third-party application servers.

Current O-RAN specifications rely primarily on IPsec to secure \ac{UP} traffic, complemented by \ac{TLS}/\ac{DTLS} for selected control and management interfaces \cite{oran_security_spec}. While this model provides a well-understood security baseline, it adds operational complexity and configuration overhead, especially in environments with many tenants and dynamic connectivity requirements. Furthermore, IPsec alone does not address all the emerging threats identified in recent O-RAN risk analyses, such as rogue gNBs, compromised edge functions, and man-in-the-middle attacks along vendor paths.

This paper explores WireGuard as a lightweight security for industrial Open RAN. The work leverages a real 5G Terafactory testbed, where external partners access factory devices through an open RAN infrastructure under realistic traffic and operational conditions. The proposed architecture uses WireGuard to encrypt user-plane traffic between \acp{UE} and either the \ac{UPF} or external application servers, and to authenticate the N2 control-plane connection between gNB and \ac{AMF}. This dual role allows WireGuard to protect data confidentiality against untrusted RAN elements while preventing unauthorized or rogue base stations from interacting with the core.

The contributions are threefold. First, a threat-driven design for integrating WireGuard with existing O-RAN security mechanisms. Second, a complete implementation in the Terafactory environment with role-based \ac{UE}, gateway, and core configurations. Third, a quantitative comparison of WireGuard and IPsec in terms of throughput, \ac{RTT}, and \ac{CPU} overhead, showing that WireGuard offers comparable performance with lower configuration complexity and broader applicability across interfaces. These results indicate that WireGuard is a strong candidate to complement IPsec as part of a unified security layer for O-RAN-based private networks.

\vspace{-2mm}
%%%%%%%%%%%%%%%%%%%%%%%%%%%%%%%%%%%    Section Background   %%%%%%%%%%%%%%%%%%%%%%%%%%%%%%%%%%%%%%

\section{Background and motivation}

Open RAN architectures introduce security challenges that extend beyond traditional cellular networks. By disaggregating RAN functions and enabling multi-vendor deployments, Open RAN exposes new interfaces, software components, and supply chains to potential attacks. Sajid et al.~\cite{wg-paperj3} survey these threats, highlighting risks from rogue base stations, compromised software modules, and untrusted intermediate network elements. In industrial campus networks, these vulnerabilities intensify when external vendors require remote access to factory equipment through shared RAN infrastructure. Although the O-RAN Alliance mandates IPsec for user-plane protection and TLS/DTLS for control-plane interfaces~\cite{oran_security_spec}, Groen et al.~\cite{secure-cost} demonstrate that IPsec introduces notable performance overhead and operational complexity in latency-sensitive scenarios.

User-plane encryption is essentially critical in industrial deployments where production data may traverse untrusted network segments. Current specifications rely primarily on IPsec, which supports multiple cipher suites, but requires complex \ac{IKEv2}-based policy negotiation. This flexibility comes at the cost of higher configuration overhead and a greater risk of misconfigured security policies. Studies of \ac{VPN} alternatives for industrial systems suggest potential improvements: Lackorzynski et al.~\cite{wg-paper5} find that lightweight VPNs can reduce latency while maintaining security, though their evaluations are not specific to Open RAN.

WireGuard has emerged as a promising lightweight alternative. Donenfeld~\cite{wg-def} designed WireGuard with a minimal cryptographic codebase ($\approx$ 4{,}000 lines) using fixed algorithms (Curve25519, ChaCha20-Poly1305), avoiding the configuration complexity of  IPsec. Performance evaluations consistently favor WireGuard: Anyam et al.~\cite{wg-paperj4} show higher throughput and lower latency than OpenVPN, and Gentile et al.~\cite{gentile2022vpn} confirm its efficiency on resource-constrained IoT hardware. In 5G scenarios, Haga et al.~\cite{wg-paper6} achieve 5.3 times higher throughput for network slice isolation, and Michaelides et al.~\cite{wg-paper2} measure only 260\,$\mu$s latency compared to 600\,$\mu$s for  IPsec. Esmaeily and Kralevska~\cite{wg-paperj} extend these concepts to multi-tenant slice orchestration.

Despite these advances, prior work remains limited to simulations and laboratory testbeds. No study has evaluated WireGuard in an operational Open RAN deployment that addresses both user-plane encryption and control-plane authentication under realistic industrial conditions. This paper addresses this open gap.

\vspace{-2mm}

%%%%%%%%%%%%%%%%%%%%%%%%%%%%%%%%%%%    Section Problem definition   %%%%%%%%%%%%%%%%%%%%%%%%%%%%%%%%%%%%%%

\section{Threat model and Security requirements \label{secreq}}

\begin{figure}[h!]
    \centering
    \includegraphics[width=\linewidth]{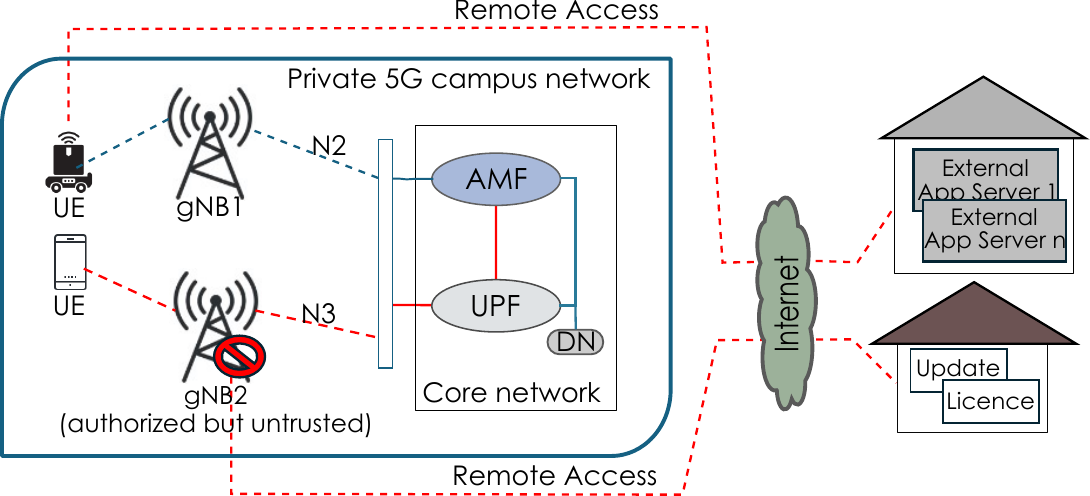}
    \vspace{-4.5mm}
    \caption{O-RAN-based private network architecture}
    \label{fig:deploy-archi}
    \vspace{-2.5mm}
\end{figure}

%\subsection{Deployment Context and Motivation}
This work builds on a series of evaluation studies conducted within our Open RAN–based private industrial 5G deployment, where multi-vendor RU, DU, and CU components were integrated and operated in compliance with O-RAN Alliance specifications. The network incorporates two gNB implementations---an open-source \ac{OAI}-based gNB and a proprietary solution---combining the advantages of both approaches. Their performance was previously evaluated in~\cite{TechRxiv25}, contributing a multi-vendor deployment experience from an operational factory environment.
%Both gNBs interface with the \ac{OAI} 5G Core, which includes full functionality (AMF, UPF, UDM, AUSF, NSSF, etc.). 
The network hosts diverse industrial applications, including robots, manufacturing stations, and industrial PCs, while external partners require remote access for updates, licensing, and maintenance. This operational topology, as depicted in Figure~\ref{fig:deploy-archi}, introduces critical security boundary crossings:

\begin{enumerate}
    \item The \ac{UP} traffic traverses the N3 interface between the gNBs and the UPF over a local network without dedicated isolation.
    \item \ac{CP} signaling is transmitted over the N2 interface between the gNBs and the AMF.
    \item Factory network connects to external service infrastructures via standard Internet links, and partner-managed devices share the same 5G connectivity footprint as internal assets.
\end{enumerate}

Given this architecture, our multi-vendor Open RAN deployment exposes potential security vulnerabilities that are not yet fully addressed by existing O-RAN Alliance specifications.
%This motivates our research into lightweight, practical security mechanisms tailored for private industrial 5G environments.

\vspace{-2.5mm}

  %%%%  Campus Open RAN deployment in the Terafactory, showing two gNB implementations, the OAI 5G Core (AMF, UPF), and remote access paths for external partners and vendor maintenance.  %%%%%%%%%   I have to refine the figure...
\vspace{-2.8mm}
\subsection{Threat Model}
We identify four primary threat scenarios associated with the deployment and operational requirements of the private 5G network.

\subsubsection{\ac{UP} Traffic Exposure via Untrusted gNB}
Although the proprietary gNB is authorized within the infrastructure, neither its code integrity nor the absence of backdoors can be assured. If compromised or controlled by an adversary, the gNB obtains direct access to decrypted user-plane data at the \ac{PDCP} layer, enabling inspection, logging, or exfiltration of complete payloads to external entities.

\subsubsection{External Access Exposure}
Remote access for updates, diagnostics, and maintenance, typically provided through VPN gateways terminating inside the factory network, introduces additional risk. Misconfigurations or insufficient access controls may permit unauthorized lateral movement, exposure of internal network services, or unintended access by third parties.
   
\subsubsection{ \ac{CP} exploitation} An untrusted gNB with backhaul connectivity may act as a rogue gNB by replicating factory network identifiers, causing \acp{UE} to attach and initiate N2 signaling with the \ac{AMF}. Without robust mutual authentication on N2, the attacker can inject forged signaling messages, disrupt registrations and handovers, enforce security downgrades, or collect UE identifiers. These risks increase when private networks delay or relax certificate-based authentication as prescribed by by 3GPP.

%%%% stopped here....30.11
\subsubsection{Inference of Internal Network Structure Through Remote-Access Abuse} If an external service account or VPN tunnel is compromised, an attacker can perform reconnaissance--such as timing analysis, port scanning, and DNS queries--to map network topology, identify core functions such as the UPF or management servers, and enumerate device roles. Even without immediate data exfiltration, such insights enable targeted technical attacks and support tailored social engineering campaigns.    

\vspace{-1mm}
\subsection{Security Requirements}
From the identified threats, we derive the following security requirements:

   \subsubsection{End-to-End \ac{UP} Confidentiality} User-plane traffic shall remain protected end-to-end, from the UE to the \ac{UPF}, without reliance on the trustworthiness of intermediate gNBs.
    \subsubsection{\ac{UP} Integrity} An untrusted or compromised gNB shall not be able to modify application payloads within user-plane traffic. Integrity protection must ensure that any alteration is detectable before delivery to the processing network function.
    \subsubsection{Network isolation} External partners shall be restricted to explicitly designated network resources. Isolation must be enforced to prevent lateral movement into internal factory subnets or sensitive infrastructure. 
    \subsubsection{Mutual Authentication} Gateways handling N2 signaling shall verify gNB identity before processing any attachment or session requests. Connection attempts from unauthorized gNBs must be rejected before \ac{NAS} message handling.    
    \subsubsection{Low Security Overhead} Security mechanisms shall not impose performance costs that violate industrial application tolerances. Configuration and key management must remain lightweight to ensure that deployment and lifecycle operations do not introduce operational constraints.
\vspace{0.4mm}

 %   \subsubsection{How can we secure the user plane against an authorized-but-untrusted gNB and counter malicious activity from external sources, using a lightweight, low-cost approach suitable for private industrial 5G? Our focus is on practical deployment inside the terafactory, prioritizing simplicity and performance over heavy, costly solutions.}

%\subsection{Problem Statement}

 %   \subsubsection{How to secure UE-to-Enterprise communication?}
  %  E2E UP encryption...
  %  we avoid to  apply Encryption many time at many locations...

   % \subsubsection{How to isolate untrusted  gNB?}
  %  Authentication between RAN elements:  DU-CU, CU-core (AMF) authentication...
  %  this block CU-AMF connection and avoid untrusted/rogue/malicious gNBs.

\vspace{-2mm}

%%%%%%%%%%%%%%%%%%%%%%%%%%%%%%%%%%%    Section Proposed solution   %%%%%%%%%%%%%%%%%%%%%%%%%%%%%%%%%%%%%%

%\section{Proposed solution: WireGuard for Open RAN Security}

\section{Proposed Lightweight Solution}
This section presents a lightweight security architecture for industrial open RAN deployments, aligned with the requirements outlined in the previous section. The proposed approach implements WireGuard to achieve two primary security functions: (i) user plane protection (N3), through end-to-end encryption, deployed in two scenarios depending on traffic destination, and (ii) mutual authentication between gNB and AMF on the N2 interface.
The performance of WireGuard is compared to IPsec in equivalent configurations for user plane encryption.  The goal is to propose a lightweight solution that is easy to manage, efficient, and minimizes configuration and operational complexity, avoids redundant encryption at multiple points, and addresses the threats defined in Section~\ref{secreq}.
%The evaluation examines operational simplicity, lifecycle management, security properties, and quantitative performance metrics.

\vspace{-0.5mm}
\subsection{WireGuard}
WireGuard is a modern, open-source VPN protocol that provides secure point-to-point encrypted and authenticated tunnels at the network layer using state-of-the-art cryptography. Unlike IPsec, which supports a wide range of configurable cipher suites, key exchange mechanisms, and authentication modes, WireGuard intentionally restricts its cryptographic choices to eliminate misconfiguration risks and reduce implementation complexity. The protocol uses \textit{Curve25519} for public-key cryptography and key exchange, \textit{ChaCha20} for symmetric encryption, \textit{Poly1305} for message authentication codes, and \textit{BLAKE2s} for hashing \cite{wg-def}\cite{wg-paper5}. This fixed cryptographic construction yields a codebase of $\approx$ 4{,}000 lines, compared to more than 100{,}000 lines in IPsec implementations, simplifying security audits and minimizing attack surface.

WireGuard operates by associating each peer with a static public key and a defined set of allowed IP addresses (crypto key routing), enabling automatic source validation and preventing unauthorized traffic injection. Tunnel establishment requires only a single round-trip handshake based on the Noise Protocol framework, achieving significantly lower connection latency compared to other protocols. 
In this work, we propose WireGuard to support the previously introduced security challenges in our deployment.
%IKEv2 negotiation, which involves multiple message exchanges, certificate validation, and policy agreement phases. 

%In contrast, IPsec relies on the  protocol for session establishment and supports multiple encryption algorithms (AES-GCM, AES-CBC, 3DES), authentication methods (RSA, ECDSA, pre-shared keys), and integrity checks (SHA-256, SHA-384, SHA-512). While this flexibility enables backward compatibility and adherence to diverse regulatory requirements, it also introduces operational complexity and potential for weak configurations, particularly when legacy ciphers or deprecated key sizes are mistakenly selected. IPsec's encapsulation (via Authentication Header or Encapsulating Security Payload) adds protocol overhead and state management complexity, whereas WireGuard's stateless design with implicit session keys reduces resource consumption and simplifies failover and roaming scenarios.

\vspace{-1.5mm}
\subsection{Two-layer solution approaches}
We used WireGuard to support two primary security functions, with function 1 (\ac{E2E} \ac{UP} encryption) is deployed in two distinct configurations depending on trust assumptions and operational requirements.

\begin{figure}[t]
    \centering
    \includegraphics[width=\linewidth]{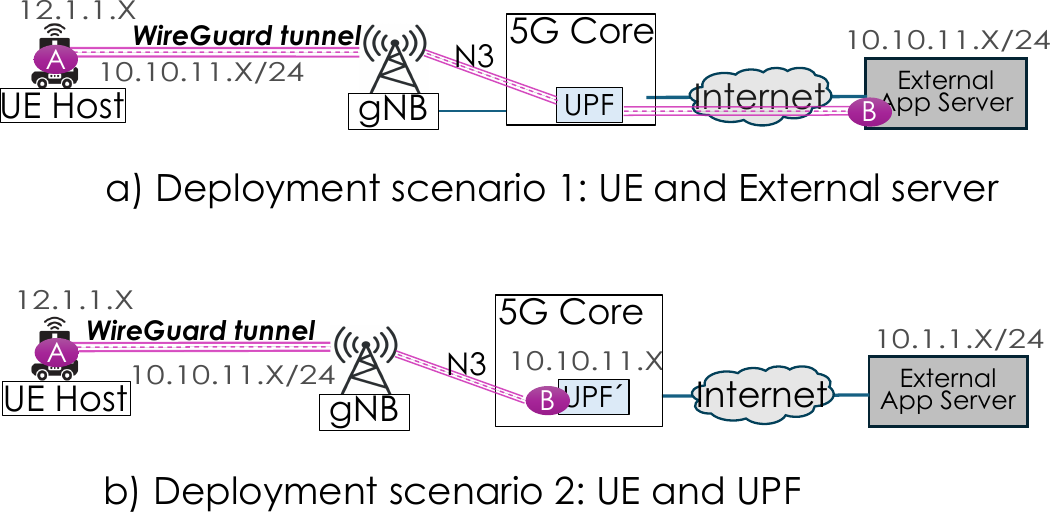}
    \vspace{-3.5mm}
    \caption{\ac{E2E} \ac{UP} Encryption on two distinct termination points: a) external server and b) UPF}
    \label{fig:e2e}
    \vspace{-4.5mm}
\end{figure}

    \subsubsection{Layer 1: End-to-End \ac{UP} Encryption (UE-UPF)} 
    This approach can be deployed in two scenarios to address different trust models.

    \begin{itemize}

    \item \textbf{Scenario 1: External connection (UE-External server)}

        Our deployment involves external partners performing remote diagnostics, firmware updates other similar tasks. WireGuard tunnels are deployed between \acp{UE} and the partner's remote server (see Figure~\ref{fig:e2e}(a)), ensuring that all user data traverses the entire network (from UE, through gNB and UPF to the external server) without intermediate inspection. WireGuard’s peer-to-peer public-key architecture provides each external server or service with a static endpoint (see Figure~\ref{fig:e2e}(a)), significantly reducing the complexity of multi-tenant VPN operation. This deployment scenario enforces strict tenant isolation; the external partners access only their explicitly authorized devices. 

     \item \textbf{Scenario 2: Internal factory traffic protection (UE-UPF)}:
   
        To protect user data from untrusted gNBs, each \ac{UE} embeds a WireGuard client configured with a unique keypair and allowed \acp{IP} strictly corresponding to its assigned address range (see Figure~\ref{fig:e2e}(b)). The UPF maintains an interface for each \ac{TEID} session, with cryptokey routing precisely mapping source \acp{IP} to authorized tunnels. The user traffic is encrypted at the \ac{UE} before transmission over the radio interface, the gNB forwards encapsulated packets without access to plaintext content, and only the UPF can decrypt the user data. This deployment provides the simplest automation path, embedding WireGuard directly into the Docker deployment, making the management of multiple UE connections straightforward. 

 \end{itemize}

\subsubsection{Layer 2: \ac{CP} Authentication}
\begin{figure}[t]
        \centering
        \includegraphics[width=\linewidth]{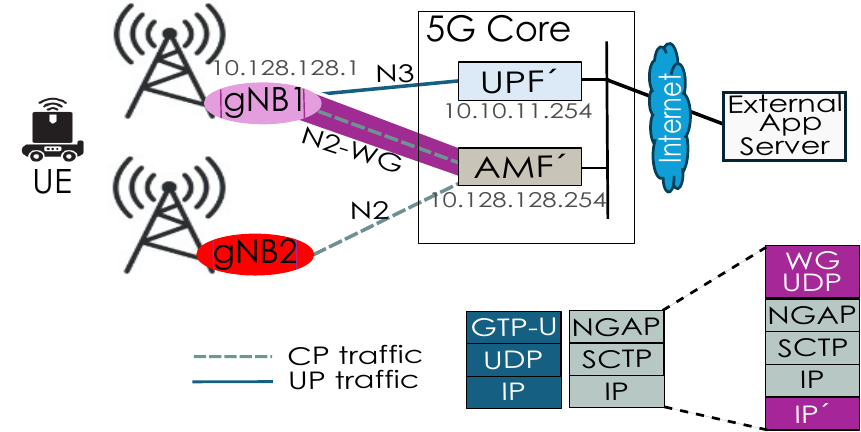}
        \vspace{-4.5mm}
        \caption{WireGuard for gNB authentication}
        \label{fig:amf}
        \vspace{-5mm}
    \end{figure}

Control plane traffic on the N2 interface between the gNB and AMF is a primary target for impersonation, signaling abuse, and rogue base station attacks. To provide Peer-to-Peer RAN element (CU and Core) authentication and mitigate these risks, a WireGuard interface is terminated on the AMF, blocking all \ac{SCTP}/\ac{NGAP} traffic from unauthenticated sources (see Figure~\ref{fig:amf}). Only gNBs presenting valid WireGuard keys can establish tunnels and reach the \ac{NGAP} layer. The gateway maintains a whitelist of authorized gNB public keys and forwards N2 signaling exclusively from authenticated peers.

This key-based authentication enables rapid revocation: a misbehaving or compromised gNB can be immediately isolated by removing its public key and reloading the configuration, preventing further control-plane communication with the core network.

%%%% Add this comments:
%% Implementing WireGuard at the UPF level is easier and more practical in our deployment because it integrates seamlessly with Docker, which is already used in our implementation. Its automation is also simpler since it does not rely on any external server and can be configured for multiple UEs simultaneously. In contrast, implementing WireGuard directly on external application servers represents the typical use case but requires access to partner equipment, making automation difficult. For each application server, WireGuard would need to be deployed separately, which would significantly increase overall complexity.

%%%%%%%%%%%%%%%%%%%%%%%%%%%%%%%%%%%%%% Section Testbed and setup  %%%%%%%%%%%%%%%%%%%%%%%%%%%%%%%%%%%%%%%5
\vspace{-1.5mm}
\section{Testbed and Implementation}

\subsection{Testbed Architecture}

    To validate our approach, we employ an operational open RAN deployed at the Terafactory, supporting industrial robotics and automated product provisioning use cases. The testbed architecture follows the deployment architecture as depicted in Figure~\ref{fig:deploy-archi} and \ref{fig:e2e}, including a 5G core network based on \ac{OAI} v2.1.9 on a Dell PowerEdge R660 server (Intel Xeon Gold 6442Y processor with 24 cores, 2.6 GHz and 128 GB RAM), with all network functions containerized using Docker. Two gNBs provide coverage: gNB1 (\ac{OAI} v2.0.0, band n78, deployed on a Dell PowerEdge R660 server) and gNB2 (proprietary, based on an Intel Xeon D-2177NT processor with 14 cores, 1.90 GHz, and 64 GB RAM, and integrates a Genevisio PAC-010 PCIe x8 inline DU accelerator card). The deployment used commercial RU from LiteOn (FlexFi FF-RF107814) and WNC R1220. The industrial UE is a robot with a Linux PC embedded with a Quectel 5G module (RM520N-GL). An external partner gateway terminates remote access tunnels over the public Internet.

\vspace{-0.7mm}

\subsection{WireGuard and IPsec Implementation}

\begin{figure}[t]
\begin{lstlisting}[style=configstyle,label={list1},caption={WireGuard configuration on UE}]
[Interface]
#PublicKey =
PrivateKey = 
Address = 10.10.11.X/24
ListenPort = 51000

[Peer] ## Scenario 1 (UE-external server)
PublicKey = 
Endpoint = 10.1.1.X:51000
AllowedIPs = 10.10.11.X/24
PersistentKeepalive = 25

[Peer] ## Scenario 2 (UE-UPF)
PublicKey = 
Endpoint = 192.168.70.X:51000
AllowedIPs = 10.10.11.X/24, 10.1.1.X/24
\end{lstlisting}
\vspace{-4.5mm}
\end{figure}

    \subsubsection{User plane protection}
Following attachment to the network, the UE receives IP address from \texttt{12.1.1.0/24} range, and establishes a WireGuard tunnel on the single interface \texttt{wg0} in the \texttt{10.10.11.X/24} listening on UDP port~51000. The configuration follows a role-based design: the UE selects its peer and \texttt{AllowedIPs} according to the deployment scenario. In Scenario~1 (UE–external server), the UE peers with the external server at \texttt{10.1.1.X:51000}, restricting \texttt{AllowedIPs} to its own address (as shown in Listing~\ref{list1}). In Scenario~2 (UE–UPF), the UE peers with the UPF at \texttt{192.168.70.X:51000} and sets \texttt{AllowedIPs} to both \texttt{10.10.11.X/24} and \texttt{10.1.1.X/24} to enable routing to internal and external services. UPF and external-server configurations (Listing~\ref{list2}) whitelist the entire UE subnet \texttt{10.10.11.0/24} and set their endpoints~\cite{wireguard}.

For comparison, an equivalent configuration was implemented using IPsec. IPsec is configured with strongSwan v5.9.8 using IKEv2, ESP tunnel mode, and 3GPP Profile~1 (AES-256-GCM with HMAC-SHA256), following the standard strongSwan installation and configuration guidelines~\cite{ipsec}.

\begin{figure}[t]
\begin{lstlisting}[style=configstyle,label={list2},caption={WireGuard configuration on UPF/external server}]
[Interface]
#PublicKey =
PrivateKey =
Address = 10.10.11.X/24
ListenPort = 51000

[Peer]
PublicKey = 
Endpoint = 12.1.1.X:51000
AllowedIPs = 10.10.11.0/24
PersistentKeepalive = 25
\end{lstlisting}
\vspace{-5mm}
\end{figure}

\begin{comment}
    
\begin{lstlisting}[style=configstyle,label={list2},caption={WireGuard configuration on UPF/external server}]
[Interface]
#PublicKey =
PrivateKey =
Address = 10.10.11.X/24
ListenPort = 51000

[Peer]
PublicKey = 
Endpoint = 12.1.1.X:51000
AllowedIPs = 10.10.11.0/24
PersistentKeepalive = 25
\end{lstlisting}

\end{comment}

    \subsubsection{Control plane authentication}

        Wireguard is additionally deployed as previously on the \ac{CU} and the \ac{AMF} to provide mutual authentication between the gNB and the \ac{AMF}, who will share the role. Only gNBs with valid public keys can establish tunnels, and all \ac{SCTP}/\ac{NGAP} packets arriving outside the tunnel are dropped. 

\vspace{-1.5mm}
\subsection{Measurement Setup}
The evaluation focuses on Scenario~1 (see Figure~\ref{fig:e2e}(a)), which corresponds to the actual deployment in our factory and reflects normal industrial operation. We assess the cost of securing the user plane with WireGuard on network \acp{KPI} and compare the results with IPsec.

Three metrics are measured: throughput using \texttt{iperf3}, \ac{RTT} latency using \texttt{ping}, and \ac{CPU} utilization. After the \ac{UE} attaches to the 5G network, we first measure these KPIs without any security mechanism to establish a baseline. We then enable WireGuard encryption and record the same metrics, followed by IPsec under identical conditions. Each configuration is tested five times to ensure consistency.
\vspace{+0.5mm}

\vspace{-2mm}

%%%%%%%%%%%%%%%%%%%%%%%%%%%%%%%%%%%    Section Evaluation   %%%%%%%%%%%%%%%%%%%%%%%%%%%%%%%%%%%%%%

\vspace{-1mm}
\section{Results and Evaluation \label{sec:results}}

This section evaluates the cost of securing the user plane in deployment scenario~1 (Figure~\ref{fig:e2e}(a)) by comparing Wireguard and IPsec. The analysis considers three \acp{KPI}: throughput, latency, and CPU utilization.

\vspace{-0.5mm}
\subsection{Latency Analysis}

\begin{figure}[t!]
    \centering
    \includegraphics[width=\linewidth]{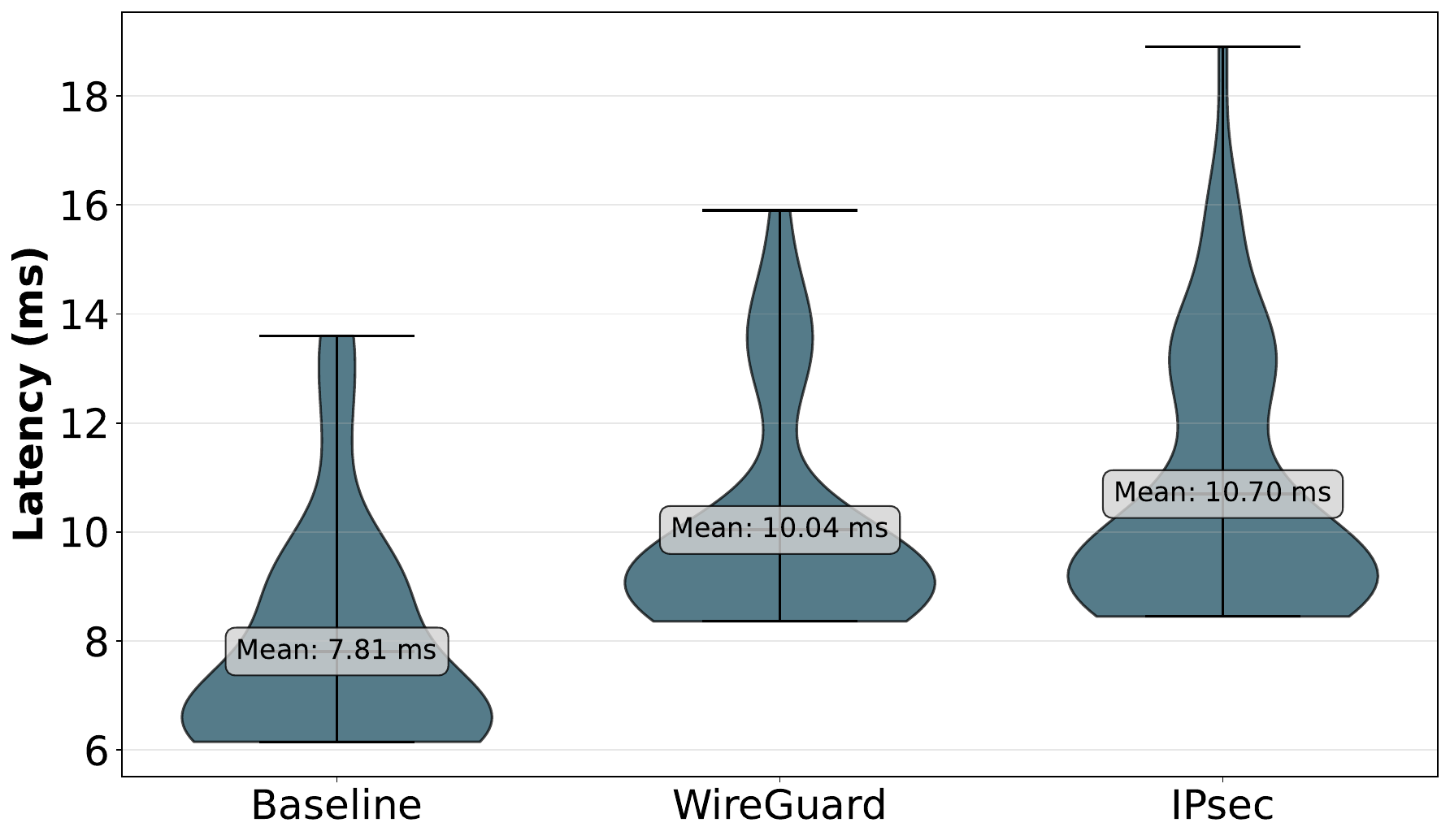}
    \vspace{-4.5mm}
    \caption{Security cost on latency and comparison with IPsec.}
    \label{fig:latency}
    \vspace{-3.5mm}
\end{figure}

RTT latency measurements are presented in Figure~\ref{fig:latency}. The baseline exhibits a mean latency of 7.81~ms with relatively low variance (95th percentile: 13.6~ms). WireGuard introduces a mean overhead of 2.23~ms, resulting in an average latency of 10.04~ms (28.5\% increase), while IPsec reaches 10.70 ms (+37\%), adding 0.66~ms over WireGuard. Both VPNs exhibit bimodal distributions, likely due to encryption-induced processing variability, whereas the baseline remains tightly concentrated. At the 99th percentile, WireGuard and IPsec record respectively 16.0~ms and 18.5~ms versus 13.7~ms for the baseline. WireGuard then demonstrates less impact on the latency while implemented. Despite overhead, both protocols remain suitable for industrial applications.
%\vspace{-3.8mm}
\subsection{Cost on Throughput}

\begin{figure}[t!]
    \centering
    \includegraphics[width=\linewidth]{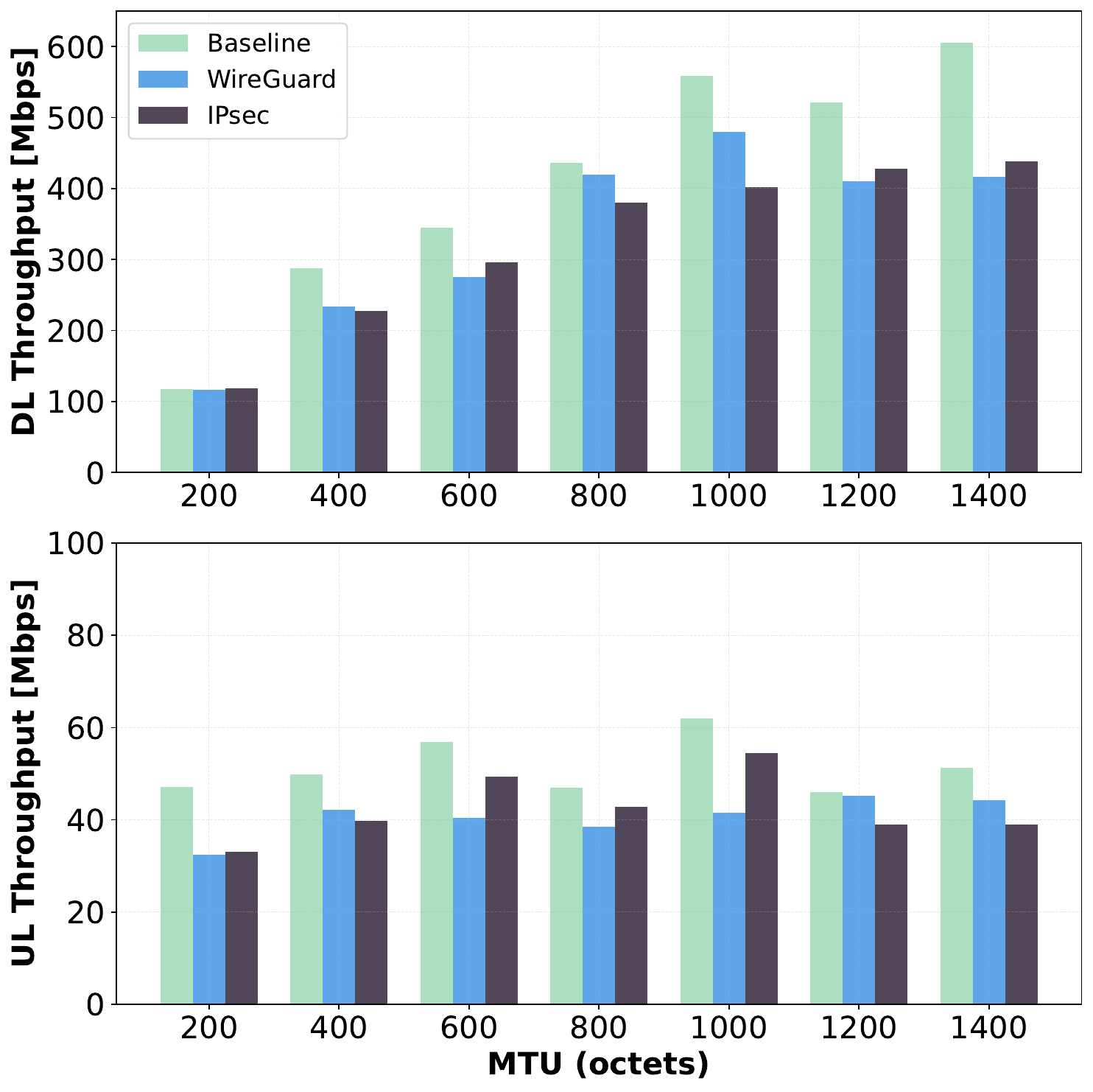}
    \vspace{-4.5mm}
    \caption{Security cost on throughput}
    \label{fig:throughput}
    \vspace{-4.3mm}
\end{figure}

Figure~\ref{fig:throughput} compares the cost of deploying WireGuard and IPsec to secure the user plane by showing downlink and uplink throughput measured across different MTU sizes. The baseline (no VPN) achieves up to 600~Mbps DL and 60~Mbps UL. WireGuard maintains near-baseline performance up to 480~Mbps DL (MTU~1000) but drops to approximately 340~Mbps at MTU~1400 (43\% reduction). IPsec exhibits more consistent behavior across all MTU ranges, outperforming WireGuard by approximately 2--5\%. Uplink trends are similar with smaller absolute values: baseline 60~Mbps, WireGuard 45~Mbps ($-25$\%), IPsec 40~Mbps ($-33$\%). 
Overall, WireGuard demonstrates consistent performance across nearly all MTUs and in both uplink and downlink directions, achieving throughput comparable to IPsec while delivering lower security overhead.

\vspace{-0.5mm}
\subsection{CPU Utilization}

\begin{figure}[t]
    \centering
    \includegraphics[width=0.85\linewidth]{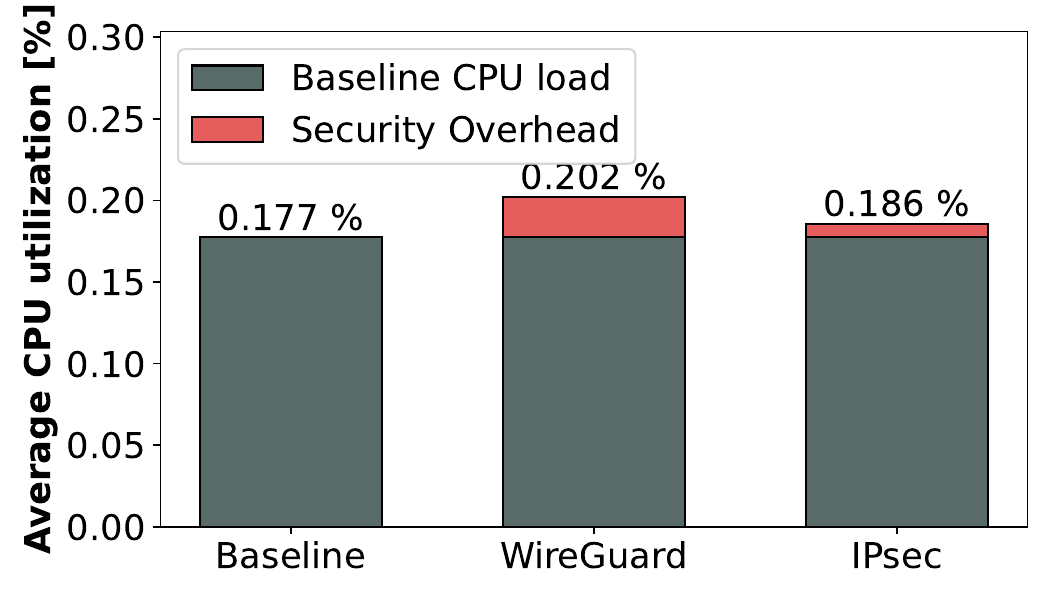}
    \vspace{-3.5mm}
    \caption{CPU cost of Encryption}
    \label{fig:cpu}
    \vspace{-4.5mm}
\end{figure}

Figure~\ref{fig:cpu} summarizes the average CPU cost of user‑plane encryption for baseline, WireGuard, and IPsec. We repeat the measurement ten times, and only the effective CPU utilization attributable to encryption was considered. Baseline exhibits 0.177\%, which increases to 0.202\% with WireGuard and 0.186\% with IPsec. The additional overheads remain well below 1\% of total CPU capacity, indicating that user‑plane encryption introduces negligible processing load and both protocols are suitable for resource‑constrained industrial edge deployments, where processing resources are limited.

\vspace{-1.5mm}
\subsection{Discussion and implications for O-RAN Security}

In the O-RAN specifications, user-plane security is ensured through IPsec. Our results show that WireGuard achieves comparable performance with lower latency overhead (2.23 ms vs. 2.89 ms, a 28.5\% reduction) and similar throughput and CPU efficiency. WireGuard's fixed cryptographic suite (Curve25519, ChaCha20-Poly1305) eliminates IPsec's configuration complexity, reducing misconfiguration risks and operational overhead. Its cryptokey routing provides implicit authentication without complex policy management, making it attractive for industrial deployments.

These findings suggest that the O-RAN Alliance should consider WireGuard as an alternative security protocol for multiple interfaces. Current specifications focus on IPsec, TLS, and DTLS, but WireGuard could address gaps identified in the O-RAN Threat Model, particularly for: (1) Open Fronthaul—low latency enables C/U/S-plane encryption without violating timing constraints; (2) F1/E2 interfaces—securing O-DU/O-CU and \ac{RIC} traffic, (3) Management plane (O1/O2)—simplified remote access, and (4) xApp/rApp communication—lightweight distributed RIC security.

The O-RAN Alliance's 2025 roadmap emphasizes closing zero-trust gaps across the architecture. In this context, WireGuard's peer-to-peer model aligns with zero-trust principles by securing each function independently. As O-RAN deployments incorporate multi-vendor scenarios and dynamic workloads, WireGuard's stateless design and single-round-trip handshake offer a significant advantage over IKEv2, enabling faster tunnel provisioning and supporting rapid enforcement of security policies.

\vspace{-2mm}

%%%%%%%%%%%%%%%%%%%%%%%%%%%%%%%%%%%    Section Conclusion   %%%%%%%%%%%%%%%%%%%%%%%%%%%%%%%%%%%%%%

\section{Conclusion and future works}

This paper evaluated WireGuard as a lightweight alternative to IPsec for securing the user plane in an industrial Open RAN deployment at the Terafactory campus. The architecture uses WireGuard to encrypt UE traffic toward the UPF or external application servers and to authenticate the N2 interface between gNB and AMF. Measurements show that WireGuard introduces moderate but acceptable overhead: latency increases by about 2.23 ms, throughput reduction remains within 18–43\% depending on MTU, and CPU time rises by roughly 3\%, comparable to IPsec. At the same time, WireGuard significantly simplifies configuration and key management. These results indicate that WireGuard is a practical candidate to complement IPsec in O-RAN-based industrial networks.

\vspace{-2mm}

\begin{comment}
    
\begin{table}[htbp]
\caption{Table Type Styles}
\begin{center}
\begin{tabular}{|c|c|c|c|}
\hline
\textbf{Table}&\multicolumn{3}{|c|}{\textbf{Table Column Head}} \\
\cline{2-4} 
\textbf{Head} & \textbf{\textit{Table column subhead}}& \textbf{\textit{Subhead}}& \textbf{\textit{Subhead}} \\
\hline
copy& More table copy$^{\mathrm{a}}$& &  \\
\hline
\multicolumn{4}{l}{$^{\mathrm{a}}$Sample of a Table footnote.}
\end{tabular}
\label{tab1}
\end{center}
\end{table}

\end{comment}
%\balance
\section*{Acknowledgment}
\vspace{-1mm}
This work has been supported in part by the German Federal Ministry of Research, Technology and Space (BMFTR) through the project 6G-Terafactory.

\vspace{-1.5mm}
\bibliographystyle{IEEEtran}
\bibliography{references}

%\balance
\end{document}